\documentclass[useAMS,usenatbib]{mn2e}
\usepackage{graphicx,natbib,times,mathptm,bm, amssymb}
\usepackage[fleqn]{amsmath}
{\newif\ifnotend
\notendtrue
\def\veclist{ABCDEFGHIJKLMNOPQRSTUVWXYZabcdefghijklmnopqrstuvwxyz.}
\def\top#1#2.{#1}
\def\tail#1#2.{#2.}
\loop\expandafter\xdef\csname v\expandafter\top\veclist\endcsname%
{{\noexpand\bm\expandafter\top\veclist}}
\edef\veclist{\expandafter\tail\veclist}
\if\veclist.\notendfalse\fi\ifnotend\repeat}
\def\pr{{\mathop{\hbox{p}}}}

\newcommand{\appropto}{\mathrel{\vcenter{
  \offinterlineskip\halign{\hfil$##$\cr
    \propto\cr\noalign{\kern2pt}\sim\cr\noalign{\kern-2pt}}}}}

\bibpunct{(}{)}{;}{a}{}{,}

\def\feh{[{\rm Fe}/{\rm H}]}

\def\E{\mathop{\hbox{E}}}

\def\norm{\mathcal{N}}

\def\ntot{N}

\def\selfunc{\mathcal{S}}
\def\hyper{\bm{\Theta}}
\def\data{\mathcal{D}}
\def\pparams{\{\vx_i\}}

\bibpunct{(}{)}{;}{a}{}{,}

\begin{document}

\bibliographystyle{mn2e}

\title[3D extinction mapping and selection effects]{Three-dimensional extinction mapping and selection effects}

\author[Sale]{S.~E.~Sale\\
Rudolf Peierls Centre for Theoretical Physics, Keble Road, Oxford OX1 3NP, UK\\}

\date{Received .........., Accepted...........}

\maketitle

\begin{abstract}

Selection effects can bedevil the inference of the properties of a population of astronomical catalogues, unavoidably biasing the observed catalogue.
This is particularly true when mapping interstellar extinction in three dimensions: more extinguished stars are fainter and so generally less likely to appear in any magnitude limited catalogue of observations.
This paper demonstrates how to account for this selection effect when mapping extinction, so that accurate and unbiased estimates of the true extinction are obtained.
We advocate couching the description of the problem explicitly as a Poisson point process, which allows the likelihoods employed to be easily and correctly normalised in such a way that accounts for the selection functions applied to construct the catalogue of observations.

\end{abstract}
\begin{keywords}
methods: statistical -- dust, extinction
\end{keywords}

\section{Introduction}\label{sec:intro}

A common task in astrophysics is to infer the characteristics of a population from observations of its members.
However, generally we do not observe all members of the population.
Moreover, those we do observe do not constitute a unbiased sample, but are selected in some systematic fashion.
Normally we refer to objects as being subject to some `selection function', which gives the probability of an object being observed.

As we do not possess observations of all objects in the population, nor those of an unbiased sample, any inference we make about the population will potentially be biased if the selection function is not accounted for.
This has been recognised for some considerable time: \cite{Malmquist_Hufnagel.1933} and \cite{Lutz_Kelker.1973} examine two particularly well known and well studied biases which arise as a consequence of selection functions.
The three dimensional mapping of extinction is subject to a similar effect \citep{Neckel_only.1966}, that if untreated will lead a significant bias in the estimation of extinction.

A well trodden approach to tackling these biases, as taken by \cite{Lutz_Kelker.1973}, follows in the spirit of \cite{Eddington_only.1913}, whereby one attempts to de-bias the results obtained using a correction factor.
However, the calculation of the correction factor relies on a solid understanding of the population, which is often not available, particularly as one might be investigating the relevant properties of the population.

An alternative approach follows from \cite{Jeffreys_only.1938}, who suggests that it is better to employ a forward modelling approach and predict what the data might look like, given the selection function and some free parameters, and then compare to the data in hand.
A Bayesian statistical approach is philosophically consistent with this paradigm, capable of including the selection function in the construction of a statistical model.

This paper will focus on the impact of selection functions when mapping interstellar extinction in three dimensions, though in approaching this focus some more general issues are covered.
Following this introduction, section~\ref{sec:Ad_intro} introduces the issue of selection functions when mapping extinction, demonstrating how simple methods for mapping extinction can be hopelessly biased.
This is followed by sections~\ref{sec:selfunc}, \ref{sec:pointp} and~\ref{sec:pp_cats} that provide the required mathematical and physical background to approach the problem of mapping extinction.
They discuss how to model selection functions (section~\ref{sec:selfunc}), introduce the Poisson point process (section~\ref{sec:pointp}) and present a discussion of how a point process formalism can be employed to describe astronomical catalogues, focusing on a treatment of selection functions (section~\ref{sec:pp_cats}).
Then, section~\ref{sec:Ad_better_ex} demonstrates how the Poisson point process based methodology can be employed to obtain unbiased estimates of the extinction distance relationship in a given direction, whilst section~\ref{sec:extn_law} discusses how selection effects can also influence observations of the extinction law and how it varies.
Finally, section~\ref{sec:close} sums up.

\section{Extinction mapping and selection functions}\label{sec:Ad_intro}

\begin{figure}
\includegraphics{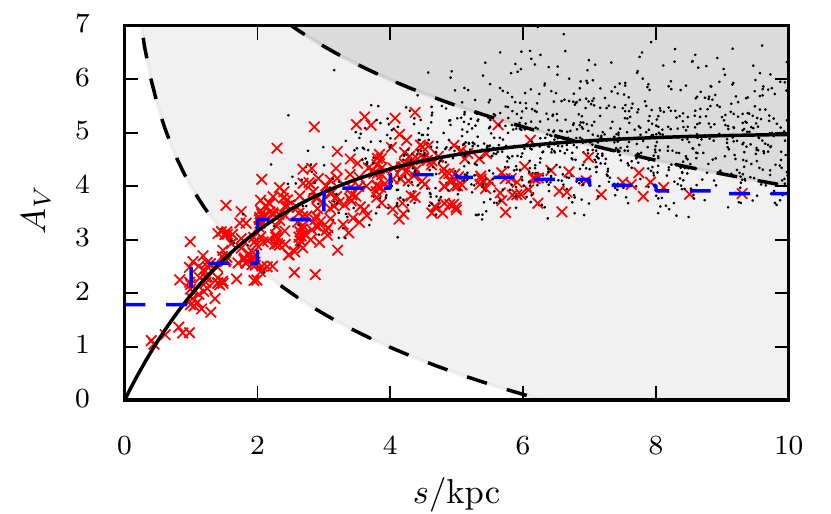}
\caption{A pseudo-photometric catalogue represented in distance--extinction space. Red crosses represent stars that appear in the catalogue, black points those that are too faint and so are not included in the catalogue.
The two black dashed lines delimit the regions where the catalogue is 100\% complete (bellow the lower line) and 100\% incomplete (above the higher line).
It can be clearly seen that the effect of the faint magnitude limit is to preferentially exclude more extinguished stars.
The mean distance-extinction relationship found by taking the mean of the extinctions of observed stars is distance bins (blue dashed line) can be compared to the true mean distance-extinction relationship (black solid line).
Not only does the binned mean underestimate extinction, but it also unphysically drops with respect to distance.
  \label{fig:extn_cat1}}
\end{figure}

In recent years there has been a growing industry in mapping interstellar extinction in three dimensions \citep[e.g][]{Marshall_Robin.2006, Sale_Drew.2009, Majewski_Zasowski.2011, Berry_Ivezic.2012, Hanson_Bailer-Jones.2014, Green_Schlafly.2014, Lallement_Vergely.2014, Schultheis_Chen.2014, Sale_Drew.2014, Sale_Magorrian.2014}.
This has been motivated both by the fact that the extinction of starlight by interstellar dust has a significant confounding impact on observations, that considerably complicates the study of individual stars and stellar populations, but also because extinction offers a direct route to studying the elusive three dimensional structure of the interstellar medium (ISM).

Three dimensional mapping of the extinction in the Galaxy is most frequently performed using photometry, owing to its relative preponderance.
As a result the data catalogues employed feature a selection function that is essentially dependent only on the sky position and apparent magnitude of the stars.
Unfortunately, the effects of the selection function are particularly pathological: more extinguished stars are fainter and thus less likely to appear in an observed catalogue, therefore a naive analysis would tend to be biased towards lower extinctions.
The existence of this problem has been recognised for some considerable time \citep[e.g.][]{Neckel_only.1966, Fitzgerald_only.1968, Neckel_Klare.1980}.

At this point, it should also be noted that photometric catalogues are typically also subject to a bright magnitude limit.
This will act to exclude the most local and least extinguished stars from a catalogue.
However, this regime is typically less well sampled due to the lower volume of space it affects.
Thus, its effects are less noticeable and so, for pedagogical purposes this paper will concentrate on the effects of the faint magnitude limit.
However, note that in practice equal care should be taken with the bright magnitude limit as with the faint.

In order to illustrate the effect of the bias caused by the faint magnitude limit, we will first consider a very simple example.
A sample of stars was simulated, with each star assigned a distance, absolute magnitude and extinction.
The details of the distance and absolute magnitude distributions are qualitatively unimportant, but follow the description given in appendix~\ref{app:Malm_ex}. 
An extinction in the $V$ band\footnote{For the sake of simplicity, it was assumed that $A_V$ depends only on the properties of the simulated ISM and that it was independent of the assumed absolute magnitudes, though note that this is not the case in reality \citep[e.g.][]{Sale_Magorrian.2015}.} was simulated assuming extinction at a given distance to be lognormally distributed \citep[following][]{Ostriker_Stone.2001}.
The mean extinction at a distance $s$ follows
\begin{equation}
\E[A_V(s)] = A_{\rm max} \left(1 - \exp(-s/2~{\rm kpc}) \right), \label{eqn:A_mean}
\end{equation}
where $A_{\rm max}$ was arbitrarily set to 5.
This mimics observing towards the Galactic anticentre.
Meanwhile, the standard deviation of extinction was defined as
\begin{equation}
\sigma[A_V(s)] = 0.3 \sqrt{\E[A_V(s)]}, \label{eqn:A_sd}
\end{equation}
inspired by \cite{Fischera_Dopita.2004} and \cite{Sale_Drew.2014}.
We further assumed that we posses perfect observations such that we know the true distance and extinctions, i.e. there are no observational errors.

In order to produce something like a real catalogue an apparent magnitude based selection function was applied to the sample: all stars brighter than a faint magnitude limit were included, whilst all those fainter were excluded.
The resultant catalogue is visualised in Fig.~\ref{fig:extn_cat1}.
As one would expect, the magnitude limit has had the effect of preferentially excluding more extinguished stars from the catalogue.
This is particularly apparent beyond $\sim 5$~kpc where almost no stars with above average extinction are observed.

One crude, though apparently appealing, way to estimate the mean extinction along a line of sight is to place stars into bins based on their distance and then find the mean extinction of stars within each bin.
Fig.~\ref{fig:extn_cat1} shows a mean distance extinction relationship estimated in this manner.
Clearly the result obtained is unsatisfactory: beyond $\sim2$~kpc the estimate is biased to smaller extinctions as a consequence of the incompleteness of the catalogue.
Furthermore, the obtained mean distance extinction relationship unphysically drops with increasing distance.
That this approach should be so unsuccessful is unsurprising given that it has ignored the role the faint magnitude limit has on shaping the observed catalogue.

There are several approaches to obtaining a better estimate of the true distance extinction relationship.
\cite{Marshall_Robin.2006}, and subsequent studies based on its method, deal with the selection function by comparing observations to  simulated data subject to a selection function that is assumed to be the same as that affecting the real data.
As a result, assuming they understand their selection function, they largely sidestep the complications provided by selection functions.
However, this approach is rigidly bound to the use of a particular Galactic model and the quality of any produced 3D extinction map will depend on the quality of the Galactic model employed.

An alternative approach, adopted by \cite{Vergely_Valette.2010}, \cite{Sale_only.2012}, \cite{Green_Schlafly.2014} and \cite{Sale_Magorrian.2014} is to build statistical models that estimate the posterior probabilities of various 3D extinction maps given some observed catalogue.
So, they need to include selection effects in the statistical model in order to calculate probabilities correctly.
In particular the selection function will affect the normalisation of the likelihood employed, since the range of observations accessible will be restricted by the selection function.
We will return to this issue in section~\ref{sec:Ad_better_ex}, where a Poisson point process based methodology will be used to infer an unbiased estimate of the distance extinction relationship in the presence of a selection function.
Prior to that, the following sections will develop the mathematical framework required in section~\ref{sec:Ad_better_ex}.

\section{Selection Functions}\label{sec:selfunc}

In general we define a function $\selfunc$ that determines the probability of a given object (potentially a star, galaxy, etc.) appearing in some catalogue. 

In all cases the selection function is conditionally independent\footnote{See appendix~\ref{app:ind} for a reminder of the definitions of conditional and marginal independence.} of the true parameters of an object given the observations on which the selection has been based.
This is trivially apparent in e.g. complete photometric catalogues that include all sources detected in an image.
However, it is not necessarily clear that it is also true in other, more involved, cases.
Consider a sample of galaxies within some redshift range compiled using photometric redshifts.
Clearly the selection will not be marginally independent of the true redshifts of the galaxies, but it is conditionally independent of them, given the photometry that has been used to derive the redshifts.
Consequently, it's possible to define a selection function that depends only on the selecting data and not on the physical parameters of the objects being studied.

Photometric catalogues are fundamentally subject to bright and faint magnitude limits: the brightest stars saturate in images and so cannot easily be included in a reliable catalogue, whilst stars below some brightness will be too faint to detect above the noise from the sky background, etc. 
Normally this part of the selection function can be estimated for entire image fields by smoothing the selection function at the limits by the use of e.g. a sigmoid function so as to account for the variable sky brightness, gain and so on across an image.
Further selections are sometimes made on photometric catalogues, e.g. colour cuts intended to select samples dominated by certain types of object.

While photometric surveys tend to shoot blind, imaging all sufficiently bright stars, spectroscopic surveys are defined by some input catalogue.
Typically the input catalogue will be compiled using data from some photometric survey.
For example APOGEE employs 2MASS \citep{Zasowski_Johnson.2013}.
Consequently the selection function will depend on the photometry employed and will be conditionally independent of the spectra obtained.
However, users will often then define subsamples to use, requiring the observed spectra to meet certain quality requirements and/or requiring \textit{measured} quantities from the spectrum (e.g. \feh) to fall within some range.
The selection function of this more restricted sample will then depend on both the photometry used to define the input catalogues and the observed spectra, including measurements derived from the spectra.

Samples selected `by hand' are generally unsuitable for use when studying a population, as it is often simply not possible to define the selection function.
An exception to this is citizen science projects, where the large number of people involved means selections made by individual humans can be treated statistically.

A general assumption that is made when considering selection functions,  that we will also make here, is to assume that the appearance of one object in a catalogue is independent of the appearance, or non-appearance, of any other objects.
As a result, we will assume that it is possible to write down a selection function for each object that depends only on the observations of that object and not on those of any other objects in the catalogue.

\section{Point Processes}\label{sec:pointp}

A point process is a random process that produces a distribution of points within some space.
Because a point process is a random process, each realisation of it will produce a distribution of points that is distinct to other realisations.
Examples of their use cover a diverse variety of situations, from the distribution of trees within some region \citep{Diggle_only.2003}, to the frequency of calls to a telephone exchange \citep{Erlang_only.1909}.

Central to our purposes here is that it is possible to consider a catalogue of astrophysical observations as a realisation of a point process in the space of observations, as we will discuss in section~\ref{sec:pp_cats}.

There are a number of different types of point process.
Here we will concentrate on one of the simplest: the Poisson point process.
In this section a short summary of only the most salient features of Poisson point processes is provided.
There are many text books available that provide a more detailed introduction to Poisson point processes and/or point processes more generally, e.g. \cite{Kingman_only.1992}, \cite{Moller_Waagepetersen.2004} and \cite{Illian_Penttinen.2008}.

\subsection{The Poisson point process}\label{sec:poisson_p}

We start by considering some region $\mathcal{Y}$ of arbitrary dimensional space, described by the position $\vy$.
Within this region we observe some distribution of points, where the points are indistinguishable except for their locations.
Then the data $\data$ consist of the set of $\ntot$ locations $\{\vy_1, \vy_2, \ldots, \vy_n \}$ where points are observed assuming no two locations are identical.

We define an intensity function $\lambda(\vy|\hyper)$ which is a deterministic function\footnote{If $\lambda$ is a stochastic, rather than deterministic function of the the hyperparameters then one has a Cox process. Cox processes are similar to Poisson point processes, but making inferences on models that employ them is somewhat complicated by their stochastic nature \citep[e.g.][]{Adams_Murray.2009}} of the position $\vy$ and a set of hyperparameters $\hyper$.
This $\lambda(\vy|\hyper)$ defines how probable it is that a point will appear at a given $\vy$.
For a given point process to be a Poisson point process we require that the number of events within some smaller region $\mathcal{Y}_S$ of $\mathcal{Y}$ is defined to be Poisson distributed with a mean of
\begin{equation}
\E[n(\mathcal{Y}_S) ] = \int_{\mathcal{Y}_S} \lambda(\vy|\hyper) d\vy
\end{equation}
so that the likelihood follows
\begin{equation}
\pr(n(\mathcal{Y}_S) | \hyper) = \frac{\left(\int_{\mathcal{Y}_S} \lambda(\vy|\hyper) d\vy \right)^n}{n!} \exp \left(-\int_{\mathcal{Y}_S} \lambda(\vy|\hyper) d\vy \right)
\end{equation}
We also require that the number of events within any two distinct regions of space are independent, conditioned upon $\lambda$.
So, for a set of disjoint regions $\{\mathcal{Y}_j\} = \{\mathcal{Y}_1, \mathcal{Y}_2, \ldots, \mathcal{Y}_M\}$, 
\begin{equation}
\pr(\data | \hyper, \{\mathcal{Y}_j\}) =  \prod_{j=1}^{M} \pr(n(\mathcal{Y}_j) | \hyper)  .
\end{equation}
A point process that satisfies these two requirements is a Poisson point process.
The likelihood for a Poisson point process is
\begin{equation}
\pr(\data | \hyper) = \exp \left(- \int_{\mathcal{Y}} \lambda(\vy|\hyper) d\vy \right) \prod_{n=1}^{\ntot} \lambda(\vy_n|\hyper) . \label{eqn:P_like}
\end{equation}
The exponential term in the likelihood is a normalisation, covering all possible number of points and all values in the space $\mathcal{Y}$ for each point.
A short derivation that demonstrates that this is a normalisation factor can be found in Appendix~\ref{app:norm}.
Critically, this normalisation depends on the hyperparameters.
So, if one seeks to infer the hyperparameters, this normalisation must be included throughout the analysis.
In some situations the integral within the exponent can be found analytically, otherwise it is necessary to find a reliable numerical approximation of it.

In many real-life situations we will be interested making some inference on the hyper parameters $\hyper$ that control the intensity function $\lambda(\vx|\hyper)$.
In order to do so we can invoke Bayes' rule, so that, in the most simple case,
\begin{equation}
\begin{split}
\pr(\hyper | \data) &= \frac{ \pr(\data | \hyper) \pr(\hyper) }{\pr(\data)} \\
& = \frac{ \pr(\data | \hyper) \pr(\hyper) }{\int \pr(\data | \hyper) \pr(\hyper) d\hyper} , 
\end{split}
\end{equation}
where the use of e.g. Markov chain Monte-Carlo (MCMC) circumvents the need to calculate the denominator.

\section{Describing catalogues as a realisation of a point process}\label{sec:pp_cats}

In what follows we are going to model astronomical catalogues as a realisation of a Poisson point process.
This is a treatment that has been explicitly adopted on a number of occasions in the past, including by \cite{Sarazin_only.1980}, \cite{Loredo_only.2004}, \cite{Gajjar_Joshi.2012}, \cite{Lombardi_Lada.2013}, \cite{Tempel_Stoica.2014} and \cite{Foreman-Mackey_Hogg.2014}.
In addition countless other authors have implicitly adopted this approach, without noting or perhaps even realising that they were doing so.
Here we will extend on these earlier works by considering the application to 3D extinction mapping including confronting the issue of selection functions.

In what follows we will assume that the space of observations includes the measured on sky position of the objects observed in addition to a range of other observations, potentially including apparent magnitudes, proper motions, parallaxes, spectra and quantities derived from the spectra.
A catalogue of data then consists of observations of these values for a list of objects.

We make four key assumptions about the data in observed catalogues:
\begin{enumerate}
\item When conditioned on the underlying `true' intensity function, the probability of an object, whether or not it is observed, existing at a point in observation space is conditionally independent of the appearance of other objects. \label{list:1}
\item The selection function does not depend on the properties of more than one object (see section~\ref{sec:selfunc}). \label{list:2}
\item The number of objects observed within some region of observation space should be Poisson distributed. \label{list:3}
\item No two observed objects can exist in the exact same location in observation space. \label{list:4}
\end{enumerate}
If the observations are made on continuous scales and conditions~\ref{list:1} and~\ref{list:2} are met, then condition~\ref{list:3} is essentially always satisfied: the law of rare events acting on arbitrarily small intervals of space allows the Poisson distribution to be adopted.
Meanwhile, the continuity of the underlying space typically ensures that condition~\ref{list:4} is met.

When satisfied, these assumptions are sufficient to meet the two conditions described in section~\ref{sec:poisson_p}, specifically the number of objects observed in any region of observation space will be Poisson distributed and the number of objects in any two disjoint regions of observation space will be independent.
Therefore, it is possible for an observed catalogue to be modelled as a realisation of a Poisson point process.

A statistical model is employed to describe the probability of objects being observed at any given point in parameter space and thus defines the intensity function $\lambda$.
The statistical model will include the selection function that acts on the data: regions favoured by the selection function will, in general, exhibit stronger intensities than those that the selection function suggest we should have trouble observing.

\subsection{Catalogue Likelihood}

In section~\ref{sec:pointp} we considered a point process intensity that was directly defined on the space of observations $\mathcal{Y}$, producing observations $\vy$.
However, in astrophysics it is more common to define physical models that set the probability of objects with some physical parameters $\vx$.
$\vx$ could include quantities such as distance, mass, metallicity and so on, and the space these physical parameters occupy is defined as $\mathcal{X}$.

We essentially never have direct and absolutely precise measurements of any physical parameter, only some noisy observations that are in some way related to $\vx$.
At best our observations may contain direct estimates of physical parameters, but more generally our observations will consist of some other quantities, such as apparent magnitudes, that depend on $\vx$ in some potentially complicated way.
The observations are related to the underlying physical parameters through a probability density function (pdf) $\pr(\vy|\vx)$, that maps from $\mathcal{X}$ to $\mathcal{Y}$, describing how likely an object with physical parameters $\vx$ is to produce observations $\vy$.

We can consider an intensity function $\lambda'(\vx, \hyper)$ that determines how frequently objects with certain physical parameters $\vx$, e.g. distance, mass and so on arise on a space of physical parameters $\mathcal{X}$, given some hyperparameters $\hyper$.
If we have a catalogue of stars, such an intensity function would include descriptions for the number density of the stars in space, an initial mass function and a star formation history.
This is then passed through $\pr(\vy|\vx)$.
Thus, if all objects were observed, we could define
\begin{equation}
\lambda(\vx, \vy | \hyper) = \pr(\vy|\vx) \lambda'(\vx | \hyper) ,
\end{equation}
where $\lambda(\vx, \vy | \hyper)$ is the intensity function on the joint space of \emph{both} the observations \emph{and} the physical parameters of the objects.

However, not all objects in the universe appear in any catalogue.
Therefore, we introduce the selection function $\selfunc(\vy)$ that describes the probability that an object with observations $\vy$ will appear in the catalogue.
So, we therefore have
\begin{equation}
\lambda(\vx, \vy | \hyper, \selfunc) = \selfunc(\vy) \pr(\vy|\vx) \lambda'(\vx, \hyper) .
\end{equation}
In many cases it is prudent to decompose the intensity function $\lambda'(\vx, \hyper)$ into the product of a normalised pdf $\pr(\vx)$ and some, possibly unknown, scaling factor $\norm$, that defines the total number of objects (observed and unobserved).
So
\begin{equation}
\lambda(\vx, \vy | \norm, \hyper, \selfunc) = \norm \selfunc(\vy) \pr(\vy|\vx) \pr(\vx |\hyper) . \label{eqn:intensity}
\end{equation}
In some cases $\norm$ may be known a priori, however, in general this will not be the case.
Note that, as it is a hyperparameter, $\norm$ could be included in $\hyper$.
However, for the sake of clarity, it is kept as a separate factor.

Now that we have defined an intensity function we can return to the formalism discussed in section~\ref{sec:pointp} and define the likelihood of some data using~\eqref{eqn:P_like}.

We consider a catalogue of data $\data$ that covers $\ntot$ objects, so that $\data = \{\vy_1, \vy_2, \ldots, \vy_{\ntot} \}$, where $\vy_n$ is the observations of the $n{\rm th}$ star.
Note that if new data were obtained, for example by re-imaging a field to produce fresh photometry, then not only will the observations of each object (i.e. $\vy_n$) change, but the size of the catalogue, $N$, may well also change, due to a different number of objects being visible in the new image.
Additionally, we define a catalogue of physical parameters of the stars $\pparams = \{\vx_1, \vx_2, \ldots, \vx_{\ntot} \}$.
Following~\eqref{eqn:P_like} and~\eqref{eqn:intensity} we get
\begin{equation}
\begin{split}
\pr(\data, \pparams| \norm, \hyper, \selfunc) &= \exp \left(-\int_{\mathcal{Y}} \int_{\mathcal{X}} d\vy' d\vx' \lambda(\vx', \vy'| \norm, \hyper, \selfunc) \right) \\
&\quad \times \prod_{n=1}^{\ntot} \lambda(\vx_n, \vy_n| \norm, \hyper, \selfunc) \\
&= \exp \left(-\int_{\mathcal{Y}} \int_{\mathcal{X}} d\vy' d\vx' \norm \selfunc(\vy') \pr(\vy'|\vx') \pr(\vx'|\hyper) \right) \\
&\quad \times \prod_{n=1}^{\ntot} \left( \norm \selfunc(\vy_n) \pr(\vy_n|\vx_n) \pr(\vx_n|\hyper) \right), \label{eqn:like}
\end{split}
\end{equation}
the likelihood for a Poisson point process on the space $\{\mathcal{X}, \mathcal{Y}\}$, the space of observations and physical parameters.
A key feature to notice is that our likelihood is conditioned on the selection function, which we must therefore know if we are to make any inference.

Often we will not be interested in the physical parameters of each of the objects $\pparams$ directly, focusing only on the hyperparameters $\hyper$ that define the population.
In that case we can opt to marginalise $\pparams$, simplifying to a Poisson point process on the space of observations only,
\begin{equation}
\begin{split}
\pr(\data| \norm, \hyper, \selfunc) &= \exp \left(-\int_{\mathcal{Y}} \int_{\mathcal{X}} d\vy' d\vx' \norm \selfunc(\vy') \pr(\vy'|\vx') \pr(\vx'|\hyper) \right) \\
&\quad \times \prod_{n=1}^{\ntot} \left( \norm \selfunc(\vy_n) \int_{\mathcal{X}} d\vx' \pr(\vy_n|\vx') \pr(\vx'|\hyper) \right) .
\end{split}
\end{equation}
Though, sometimes, it will be difficult to perform the marginalisation directly.
Therefore, it may be easier to consider the unmarginalised likelihood~\eqref{eqn:like} and use a numerical scheme, such as MCMC to marginalise $\pparams$.

\subsubsection{Known size catalogues}\label{sec:known}

We now consider catalogues of an a priori known size.
We are therefore considering a Poisson point process conditioned on the total number of observations.
Such a point process is referred to as a binomial (point) process.

Such catalogues are particularly typical of spectroscopic observations; known size catalogues typically arise when an input catalogue has been employed to produce a list of objects to be observed.

We can derive the likelihood for a fixed size catalogue by starting from that of an unknown sized catalogue, as given by~\eqref{eqn:like} and conditioning on the total number of objects $\ntot$ in the catalogue.
\begin{equation}
\pr(\data, \pparams| \ntot, \norm, \hyper, \selfunc) = \frac{\pr(\data, \pparams, \ntot| \norm, \hyper, \selfunc)  }{\pr(\ntot| \norm, \hyper, \selfunc) }  .
\end{equation}
As $\ntot$ is simply the size of the data catalogue $\data$ and the set of physical parameters, the numerator above is the likelihood given by~\eqref{eqn:like}.
$\ntot$ itself follows a Poisson distribution, with an expected number of observed stars of
\begin{equation}
\E[\ntot| \norm, \hyper, \selfunc] = \int_\mathcal{X}\int_\mathcal{Y} d\vx' d\vy' \norm \selfunc(\vy') \pr(\vy'|\vx') \pr(\vx' |\hyper).
\end{equation}
Therefore, we can say
\begin{equation}
\begin{split}
\pr(\ntot | \norm, \hyper, \selfunc) =& \frac{1}{\ntot!} \left(\int_\mathcal{X}\int_\mathcal{Y} d\vx' d\vy' \norm \selfunc(\vy') \pr(\vy'|\vx') \pr(\vx' |\hyper) \right)^{\ntot} \\
&\times \exp\left(-\int_\mathcal{X}\int_\mathcal{Y} d\vx' d\vy' \norm \selfunc(\vy') \pr(\vy'|\vx') \pr(\vx' |\hyper) \right) . \label{eqn:ntot_like}
\end{split}
\end{equation}
We can then divide~\eqref{eqn:like} by~\eqref{eqn:ntot_like} to obtain
\begin{equation}
\pr(\data, \pparams| \ntot, \hyper, \selfunc) = \ntot! \prod_{n=1}^{\ntot}  \frac{ \selfunc(\vy_n) \pr(\vy_n|\vx_n) \pr(\vx_n|\hyper) }{\int_\mathcal{X}\int_\mathcal{Y} d\vx' d\vy'  \selfunc(\vy') \pr(\vy'|\vx') \pr(\vx' |\hyper)  } . \label{eqn:like_known}
\end{equation}
Here the product on the right hand side simply corresponds to that of $\ntot$ independent draws from a normalised probability distribution proportional to $ \selfunc(\vy_n) \pr(\vy_n|\vx_n) \pr(\vx_n|\hyper) $, that includes the selection function.
Consequently it is far more intuitive than that for variable sized catalogues.
More specifically, \eqref{eqn:like_known} is simply the multinomial distribution in the limit of an infinite number of categories, where the category occupancy is never higher than 1.

This normalisation also makes sense in light of appendix~\ref{app:norm}, it is simply that acquired by integrating over all of $(\mathcal{X}, \mathcal{Y})$, but for a fixed number of objects in the catalogue, as would be expected given this probability is conditioned on the size of the catalogue.

This result is similar to that obtained by others who have worked with fixed size catalogues such as \cite{Sanders_Binney.2015}.
Note that the likelihood obtained is independent of $\norm$ and so there is no need to infer or marginalise it out.

\subsection{Prior and posterior distributions}\label{sec:prior_post}

Given that we seek to infer the hyperparameters, we now need to define the posterior probability distribution.
For the case where the size of the catalogue is fixed a priori, following~\eqref{eqn:like_known}, the posterior is:
\begin{equation}
\begin{split}
\pr(\hyper, \pparams | \data, \ntot, \selfunc) =& \frac{\pr(\data, \pparams| \ntot, \hyper, \selfunc) \pr(\hyper)}{\pr(\data|\ntot, \selfunc)} \\
=& \frac{\ntot! \, \pr(\hyper)}{\pr(\data|\ntot, \selfunc)}  \prod_{n=1}^{\ntot}  \frac{ \selfunc(\vy_n) \pr(\vy_n|\vx_n) \pr(\vx_n|\hyper) }{\int_\mathcal{X}\int_\mathcal{Y} d\vx' d\vy'  \selfunc(\vy') \pr(\vy'|\vx') \pr(\vx' |\hyper)  } . \label{eqn:posterior_known}
\end{split}
\end{equation}
If the size of the catalogue is not fixed, as is the case with photometric catalogues, following~\eqref{eqn:like}, the posterior is
\begin{equation}
\begin{split}
\pr(\hyper, \norm, \pparams | \data, \selfunc) =& \frac{\pr(\data, \pparams| \norm, \hyper, \selfunc) \pr(\hyper, \norm)}{\pr(\data|\selfunc)} \\
=& \exp \left(-\int_{\mathcal{Y}} \int_{\mathcal{X}} d\vy' d\vx' \norm \selfunc(\vy') \pr(\vy'|\vx') \pr(\vx'|\hyper) \right) \\
& \times \frac{P(\hyper,\norm)}{\pr(\data|\selfunc)} \prod_{n=1}^{\ntot} \left( \norm \selfunc(\vy_n) \pr(\vy_n|\vx_n) \pr(\vx_n|\hyper) \right)  . \label{eqn:posterior}
\end{split}
\end{equation}
We now need to place a prior on $\hyper$ and $\norm$.
Often one will assume that the priors on $\hyper$ and $\norm$ are independent, so that
\begin{equation}
\begin{split}
\pr(\hyper,\norm) &= \pr(\hyper)\pr(\norm) .
\end{split}
\end{equation}
The choice of $\pr(\hyper)$ will very much depend on the situation being studied.
In some cases a sensible $\pr(\norm)$ will also be implied.
If not, three natural choices for $\pr(\norm)$ are: a uniform prior
\begin{align}
\pr(\norm) = k, \label{eqn:uniform_prior}
\end{align}
where k is a constant; Jeffreys prior, 
\begin{align}
\pr(\norm) = \norm^{-\frac{1}{2}}; \label{eqn:Jeffreys_prior}
\end{align}
and a gamma distribution prior, which is the conjugate prior for the Poisson distribution,
\begin{align}
\pr(\norm) = \frac{\beta^{\alpha}}{\Gamma(\alpha)} \norm^{\alpha-1} e^{-\beta \norm} .
\end{align}
Notice that the uniform and Jeffreys priors are approached as special cases of the gamma distribution when $\beta \to 0$, and $\alpha=1$ or $1/2$ respectively.
Additionally, both the uniform and Jeffreys priors are improper, that is their integrals over all $\norm$ are infinite.
Improper priors often lead to improper posteriors and so should generally be avoided if possible.
Consequently, from this selection, the use of the gamma distribution prior, with $\beta \ne 0$, should be preferred.

Once armed with a fully defined posterior one can make some inference about the hyperparameters.
In general, the posteriors given by~\eqref{eqn:posterior_known} or~\eqref{eqn:posterior} will not be analytic and so one will have to employ some numerical means, such as importance sampling, MCMC or sequential Monte Carlo \citep{Doucet_DeFreitas.2001} to sample from the posterior.
In the case of a priori unknown sized catalogues, as given by~\eqref{eqn:posterior}, one has the choice of either inferring $\norm$, as it may be a parameter of interest, or marginalising over it.
Providing $\pr(\norm)$ follows a gamma distribution, we can straightforwardly perform the marginalisation to obtain
\begin{equation}
\begin{split}
\pr(\hyper, \pparams | \data, \selfunc) &= \int d\norm \pr(\hyper, \norm, \pparams | \data, \selfunc) \\
&=  \left(\beta+\int_{\mathcal{Y}} \int_{\mathcal{X}} d\vy' d\vx' \selfunc(\vy') \pr(\vy'|\vx') \pr(\vx'|\hyper) \right)^{-(\alpha+\ntot)} \\
&\quad \times \frac{\beta^{\alpha}\Gamma(\alpha+\ntot)}{\Gamma(\alpha)} \frac{\pr(\hyper)}{\pr(\data|\selfunc)} \prod_{n=1}^{\ntot}\left( \selfunc(\vy_n) \pr(\vy_n|\vx_n) \pr(\vx_n|\hyper) \right) . \label{eqn:posterior_margN}
\end{split}
\end{equation}
The marginalisation for the uniform and Jeffreys priors can be recovered by simply considering the limiting behaviour when $\beta \to 0$, and $\alpha=1$ or $1/2$ respectively, as discussed above.
Also, as $\beta \to 0$ \textit{and} $\alpha \to 0$ \eqref{eqn:posterior_margN} approaches the form of~\eqref{eqn:posterior_known}, the posterior for a fixed size catalogue.
However, as the prior approaches any of these three special cases the evidence $\pr(\data|\selfunc, \pr(\norm))$ tends to 0.

\subsection{Example: Malmquist bias}\label{sec:Malm_ex}

\begin{figure}
\includegraphics{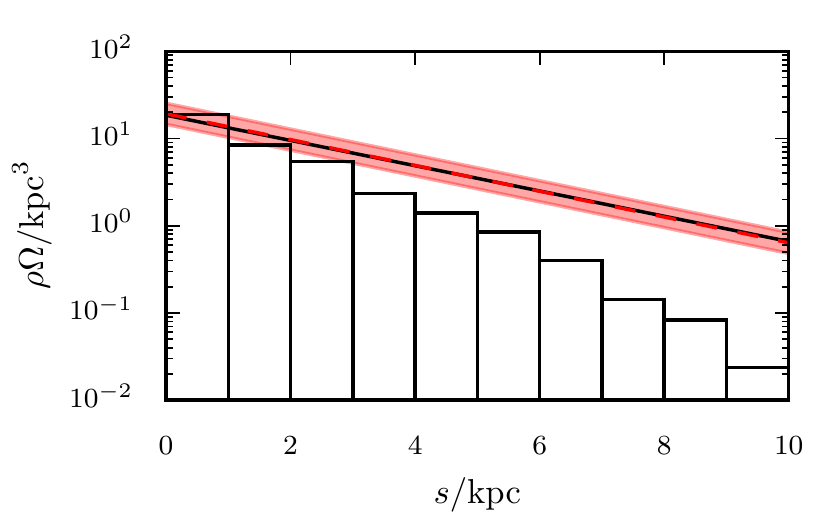}
\caption{Representations of the density of stars as a function of distance for the catalogue of stars represented in Fig.~\ref{fig:sim_cat}. The black line shows the true density, whilst the histogram shows the number of stars in the catalogue.
The dashed red line shows the posterior expectation, with the shaded area representing a 1-$\sigma$ credible interval.
Note that the posterior expectation is in near exact agreement with the true density, so that the two lines are essentially overplotted.
  \label{fig:malm_hist}}
\end{figure}

\begin{figure}
\includegraphics{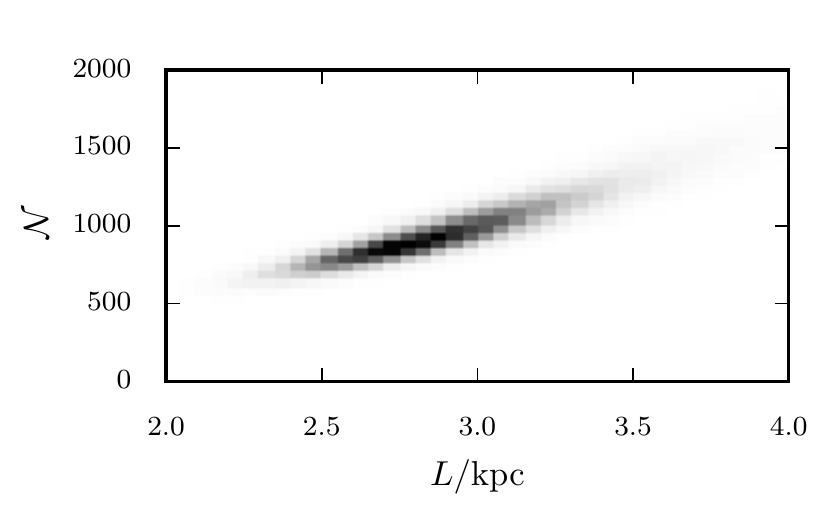}
\caption{The posterior distribution on scale length $L$ and total number of stars $\mathcal{N}$.
The true values are $L=3$~kpc and $\mathcal{N}=1000$.
  \label{fig:malm_posterior}}
\end{figure}

To initially illustrate the benefits of the point process based methodology described in the previous sections we will consider a simplified example, motivated by \cite{Malmquist_Hufnagel.1933}.
In this example, we have measured the distances to a catalogue of stars without error and now seek to estimate the distribution of these objects in space.
However, the catalogue has been subject to a faint magnitude limit and so is an incomplete and biased sample from the wider population.
The finer details of this example are discussed in appendix~\ref{app:Malm_ex}.

Given the catalogue, we now seek to infer the scale length of the population.
The most straightforward way to approach this is simply to count the number of stars in a distance bin and divide by the volume to obtain an estimate of the density, as shown in Fig~\ref{fig:malm_hist}. 
However, as expected, this approach strongly underestimates the scale length: as the distance increases the catalogue becomes progressively more incomplete, underestimating the density by more an order of magnitude beyond $\sim8$~kpc.
The scale length estimated directly from these binned counts is $\sim 1.5$~kpc, roughly half the true value.

Therefore, we apply the Poisson point process based methodology discussed in previous sections.
Using this approach we obtained a posterior distribution on the scale length and total number of stars (Fig.~\ref{fig:malm_posterior}) that agrees well with the `true' values used to simulate the data.
The density distribution implied by the posterior expectation values of both parameters is shown in Fig.~\ref{fig:malm_hist}, agreeing well with the `true' density distribution, in contrast to the naive approach.

\section{Extinction mapping}\label{sec:Ad_better_ex}

We will now return to the issue of extinction mapping and consider how to apply the Poisson point process based methodology described in the previous sections to a somewhat more realistic example than that described in section~\ref{sec:Ad_intro}.
This example is designed to largely mimic the methods employed by \cite{Sale_only.2012} and \cite{Green_Schlafly.2014}, but a number of assumptions will be made so as to reduce the complexity somewhat.

\subsection{Catalogue simulation}

A sample of 10000 stars was simulated, with the spatial distribution following that described in section~\ref{app:Malm_ex} and the extinction distribution the same as in section~\ref{sec:Ad_intro}, but for $A_r$.
In the name of simplicity all stars were assumed to exhibit solar abundances, lie on the zero age main sequence (ZAMS) as defined by \cite{Marigo_Girardi.2008} and be in single star systems (i.e. no binaries).
The mass of each star was assumed to be drawn from a power law IMF with slope -2.3, truncated to the mass range given by the \cite{Marigo_Girardi.2008} ZAMS isochrone.
In addition, the ratio of extinction between each of the bands was assumed to be constant\footnote{Note again that this is not true in practice \citep[e.g.][]{Sale_Magorrian.2015}}, set by the values for an A0V star in the limit of small extinction subject to a \cite{Fitzpatrick_only.2004} `$R_V=3.1$' reddening law.
Consequently each star is described by a distance, extinction and mass only, so that for the $n{\rm th}$ star the fundamental parameters are given by $\vx_n = \{s_n, A_{r,n}, \mathcal{M}_n\}$, the distance, extinction to and mass of the star respectively.

Given the mass, absolute magnitudes $M_X$ in each of the SDSS bands were taken from the \cite{Marigo_Girardi.2008} ZAMS isochrone.
Then apparent magnitudes in each band were simulated following
\begin{align}
\overline{m}_X &= M_X + A_r \frac{A_X}{A_r} + 5\log_{10} (s/10~{\rm pc}) \\
\sigma_X(\overline{m}_X) &= 0.02 + \exp\left( \ln(0.1) + \frac{\overline{m}_X -m_X^{\rm lim}}{2} \right)  \label{eqn:photo_uncert}\\
m_X & \sim \mathcal{N}(\overline{m}_X, \sigma_X(\overline{m}_X) ) ,
\end{align}
where $\overline{m}_X$ is the expected apparent magnitude of the star in band $X$, $m_X$ the `observed' apparent magnitude  in that band, $m_X^{\rm lim}$ the magnitude limit in this band and $\mathcal{N}$ represents the normal distribution.
The definition of $\sigma_X(\overline{m}_X)$ includes an additional factor of 0.02 to simulate systematic uncertainties, such as zero point error.
It has has been defined assuming that catalogue has a `10-$\sigma$' magnitude limit.

Then a simple magnitude limit is applied.
For each star the apparent magnitudes in each band are included in the catalogue if they are brighter than the faint magnitude limit.
If a star is too faint in all bands then it does not appear in the catalogue.
The limiting magnitudes were set to
\begin{align}
m_u^{\rm lim} &= 22\\
m_g^{\rm lim} &= 22\\
m_r^{\rm lim} &= 22\\
m_i^{\rm lim} &= 21\\
m_z^{\rm lim} &= 20.5 ,
\end{align}
roughly the typical values for SDSS.
Therefore, the selection function in band $X$ is
\begin{equation}
\selfunc^{(X)}(m_X) = \begin{cases} 1 &\mbox{if } m_X \leq m_X^{\rm lim} ;\\
0 &\mbox{otherwise. } \end{cases} \label{eqn:ad_band_selfunc}
\end{equation}
Note that in reality catalogues tend to become incomplete gradually and a sigmoid function would be a more appropriate than the step function used here \citep[e.g.][]{Green_Schlafly.2014}.

The observations $\vy_n$ for each star then comprise of apparent magnitudes in the $u$, $g$, $r$, $i$ and $z$ bands and the accompanying uncertainties given by the $\sigma_X$ from~\eqref{eqn:photo_uncert}, where observations for one or more bands may be missing.
Collectively the observations for all these stars form a catalogue $\data = \{\vy_0, \ldots \vy_{\ntot}\}$.
From the original 10000 stars simulated typically only $\sim 1000$ were bright enough to appear in the simulated catalogue.

The 3D mapping approaches of e.g. \cite{Sale_only.2012} and \cite{Green_Schlafly.2014} break the sky down into many small pixels or `resolution elements' which are then treated independently.
The size of the catalogues simulated here is not atypical of those for each resolution element in \cite{Sale_Drew.2014} and so the example we consider is a reasonable approximation to real 3D extinction mapping.

\subsection{Poisson point process based inference}

\begin{figure}
\includegraphics{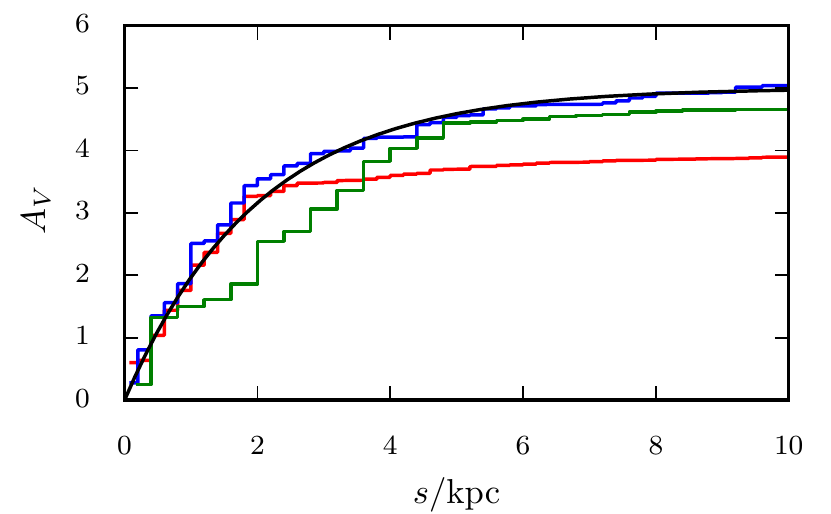}
\caption{A comparison of a mean distance-extinction relationship found by various different approaches. The true distance extinction relationship (as used to simulate the data) is shown with a black line. The red line indicates the relationship found by a \protect\cite{Sale_only.2012} based algorithm with no treatment for the selection effect, whilst the blue line shows the result obtained if a Poisson point process based treatment for the selection effects is included, with a gamma distribution prior on $\norm$. The green line indicates the result obtained by an algorithm based on the description given in \protect\cite{Green_Schlafly.2014}. 
  \label{fig:extn_better}}
\end{figure}

\begin{figure}
\includegraphics{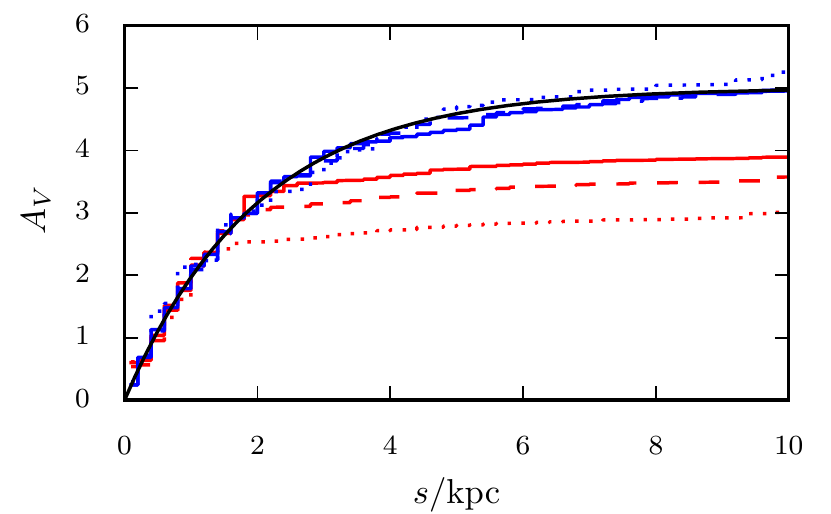}
\caption{A demonstration of the impact of raising the magnitude limit on the obtained distance-extinction relationship. The true distance extinction relationship (as used to simulate the data) is shown with a black line. The red lines show the relationships found using a \protect\cite{Sale_only.2012} based algorithm with no treatment for the selection effect, whilst the blue line shows the result obtained if a Poisson point process based treatment for the selection effects is included. In both cases the solid lines indicate a $r$ band faint magnitude limit of $m_r^{\rm lim} = 22$, dashed lines indicate a limit of $m_r^{\rm lim} = 21$ and dotted lines $m_r^{\rm lim} = 20$.
  \label{fig:extn_changed}}
\end{figure}

The point process methodology was again used to determine the characteristics of the population that $\data$ describe, specifically the dependence of extinction on distance.
The combination of the likelihood and selection function is 
\begin{equation}
\begin{split}
\selfunc(\vy) \pr(\vy_n|\vx_n) =& \Bigg[ \prod_{(m_X, \sigma_x) \in \vy_n} \frac{\selfunc^{(X)}(\overline{m}_X(\vx_n))}{\sqrt{2\pi \sigma_X^2}} \exp \left( - \frac{(m_X - \overline{m}_X(\vx_n))^2}{2 \sigma_X^2} \right) \Bigg] \\
& \times \Bigg[  \prod_{(m_X, \sigma_x) \notin \vy_n} \int \Bigg( \frac{1-\selfunc^{(X)}(\overline{m}_X(\vx_n))}{\sqrt{2\pi \sigma_X^2(\overline{m}_X(\vx_n))}}  \\
& \quad \exp \left( - \frac{(m_X - \overline{m}_X(\vx_n))^2}{2 \sigma_X^2(\overline{m}_X(\vx_n))} \right) \, d\overline{m}_X(\vx_n) \Bigg) \Bigg] , \label{eqn:Ad_obslike}
\end{split}
\end{equation}
where $\overline{m}_X(\vx_n)$ is the expected apparent magnitude of a star in band $X$ given some stellar parameters $\vx_n$.
The first line in this equation takes the product over those photometric bands for which observations exist, while the second takes the product over those bands without observations and represents the likelihood of not getting some observed photometry in that band given stellar parameters $\vx_n$.

For the purposes of this test we assume that the way stars are distributed in space, i.e.~\eqref{eqn:Malm_dist}, and the IMF are known.
Consequently the only hyperparameters are those that trace how extinction varies with distance.
Here the relationship between mean extinction and distance is parametrized with a piece-wise constant function, as in \cite{Sale_only.2012} and \cite{Green_Schlafly.2014}.
The length of each piece was set to 200~pc, in effect creating a series of 200~pc `bins'.
The standard deviation of extinction as a function of distance is described similarly.
Therefore, extinction would be described by a set of hyperparameters that are the mean ($\theta$) and standard deviation ($\zeta$) of extinction in a series of distance bins, $\hyper = \{ \theta_1, \zeta_1 , \ldots, \theta_n, \zeta_n \}$.
So, 
\begin{equation}
\pr(\vx|\hyper) = \mathcal{LN}(A_{r} | \theta_j, \zeta_j) \frac{\mathcal{M}^{-2.3}}{\int \mathcal{M}'^{-2.3} \, d\mathcal{M}'} \frac {s^2 e^{-s/(3~{\rm kpc})}}{2(3~{\rm kpc})^3} , \label{eqn:Ad_pp_int}
\end{equation}
where $\mathcal{LN}(\cdot | \cdot, \cdot)$ is the lognormal pdf defined by a mean and standard deviation.

Collectively~\eqref{eqn:Ad_obslike} and~\eqref{eqn:Ad_pp_int} define the intensity function, as in~\eqref{eqn:intensity},
\begin{equation}
\lambda(\vx, \vy | \hyper) = \norm \selfunc(\vy) \pr(\vy|\vx) \pr(\vx|\hyper),
\end{equation}
where, as before, $\norm$ represents the total number of stars that could be included in the catalogue.
This intensity function can then be used to define the likelihood following~\eqref{eqn:like}.

In order to define a posterior distribution a prior distribution $\pr(\hyper, \norm)$ must be defined.
We required that mean extinction must be positive and increase with distance and that the standard deviation of extinction must always be positive by setting the prior probability to zero if any of these conditions are broken.
Otherwise the prior on $\hyper$ was assumed to be flat and independent of $\norm$.
We will consider three possible priors on $\norm$: a uniform prior, Jeffreys prior and a gamma distribution prior, as in section~\ref{sec:prior_post}.
The gamma distribution prior assumed a mean value of 10000, with $25\%$ uncertainty, so that $\alpha=16$ and $\beta=1/625$.

As before, the simulated data are meant to represent photometry and so the the number of observed stars in the catalogue was not fixed a priori. 
Consequently, the approach discussed in section~\ref{sec:prior_post} applies and $\norm$ was marginalised following~\eqref{eqn:posterior_margN}.

A Metropolis within Gibbs MCMC scheme was employed to sample from the posterior \citep{Tierney_only.1994}: each $\vx_n$ was updated in turn, then $\hyper$ was updated in one go within each iteration of the MCMC algorithm.
The integral in the normalisation term in~\eqref{eqn:posterior_margN} was estimated numerically by passing the isochrone employed through the selection function at a range of distances along the line of sight.

Fig.~\ref{fig:extn_better} compares the posterior expectations of mean extinction against distance found as described above to the true relationship.
As is readily apparent the result obtained without any treatment for the selection function performs poorly, underestimating extinction significantly beyond 2~kpc.
The reason for this is, as described in section~\ref{sec:Ad_intro}, the observed catalogue of stars is deficient in more extinguished stars at larger distances, since they are too faint to observe.
By contrast, the posterior expectation found employing the Poisson point process methodology and a gamma distribution prior performs excellently, obtaining an accurate estimation of the true distance extinction relationship at all distances.
This approach is essentially that adopted by \cite{Sale_Drew.2014}, though simplified slightly to match assumptions made when simulating the data.

In Fig.~\ref{fig:extn_changed} we consider the impact of raising the magnitude limit used to compile the catalogue, requiring that  stars pass faint limits in the $r$ band of $m_r^{\rm lim} = 22, 21, 20$.
As the magnitude limit is raised, the results obtained using the method of \cite{Sale_only.2012} become progressively more inaccurate and depart from the true distance extinction relationship more locally as more distant and/or more extinguished stars drop out of the catalogue.
In contrast, the result obtained with the Poisson point process methodology continues to match the true distance extinction relationship, though the precision of the result decreases somewhat due to the catalogue reducing in size and so providing less information. 

Not shown in Fig.~\ref{fig:extn_better} are results found using the Poisson point process methodology and a flat or Jeffreys prior on $\norm$, as given by \eqref{eqn:uniform_prior} and \eqref{eqn:Jeffreys_prior}.
These largely match the result using the gamma function prior out to $\sim 5$~kpc at which point they unphysically diverge towards very large extinctions.

This behaviour essentially stems from the fact that both the uniform and Jeffreys priors are improper.
One consequence of employing an improper prior is that the resulting posteriors are also improper.
MCMC algorithms are unable to converge when employing an improper posterior and indeed closer examination reveals that in both cases the MCMC chains have failed to converge, with extinctions at large distances increasing to ever larger values as the chains progress.
In addition, the posteriors following from either of these priors will assign infinite probability mass to infinite values of $\norm$, which in turn allows infinite extinctions at distances beyond any observations.
Therefore, neither the uniform nor Jeffreys prior on $\norm$ should be used, and so, of the initial three options considered, only the gamma distribution prior produces realistic results.

Fig.~\ref{fig:extn_better} also displays a result based on the method described by \cite{Green_Schlafly.2014}, though again featuring simplifications to match the simulations.
In this instance this method tends to underestimate extinction, both at large distances, where the catalogue is more strongly biased towards smaller extinctions, and more locally. 
The formulation of \cite{Green_Schlafly.2014}, though similar in spirit, exhibits some fundamental differences to the approach described in \cite{Sale_only.2012} and \cite{Sale_Drew.2014}.
The most obvious difference is that they assume that differential extinction\footnote{Here differential extinction is defined as the variation in extinction with respect to changing Galactic coordinates at some fixed distance.} within any $6 \arcmin \times 6\arcmin$ region of the sky is negligible.
However, \cite{Sale_Drew.2014} demonstrates that this is not the case.
As a result of excluding small scale differential extinction, the \cite{Green_Schlafly.2014} method tends to pass very close to the posterior expectations of distance and extinction for the brightest stars, even when in reality these stars exhibit extinctions that are significantly above or below the mean value for their distance.
In this case, two bright and relatively unextinguished stars drag the \cite{Green_Schlafly.2014} estimate towards an underestimate of extinction around 2~kpc, whilst a bright star at $\sim 5$~kpc pulls the estimated mean extinction towards the true value.
However, it is unclear how, more generally, this effect will influence their results and how it will interact with their treatment for the survey selection and incompleteness.

Beyond their treatment of differential extinction, \cite{Green_Schlafly.2014} present a description of extinction is essentially the same as \cite{Sale_only.2012}: a hierarchical model under which one has some (imperfect) observations of stars that sample a distance extinction relationship.
However, there are some subtle, but fundamental, differences between the posterior distribution described in \cite{Green_Schlafly.2014} and that described here, associated with the treatment of the selection function.
A detailed discussion of the methodology of \cite{Green_Schlafly.2014}, examining how these discrepancies arise can be found in appendix~\ref{app:green}. 

\section{Extinction law variations}\label{sec:extn_law}

\begin{figure}
\includegraphics{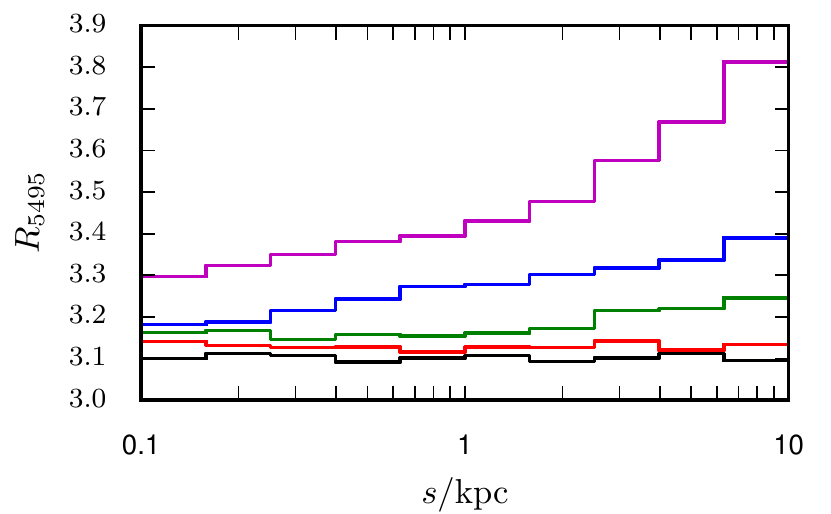}
\caption{ The mean value of $R_{5495}$ measured from stars appearing in an observed catalogue as a function of distance. The different lines indicate different values of $A_z$ applied, from bottom to top: $A_z=0.1$ (black), $A_z=0.5$ (red), $A_z=1.0$ (green), $A_z=1.5$ (blue), $A_z=2.0$ (magenta).
  \label{fig:R5495_trend}}
\end{figure}

In the previous section we examined the role selection effects can have on determining the mean extinction.
In addition, one might also be interested in the wavelength dependence of the opacity of the ISM, i.e. the shape of the extinction law, typically parametrized by $R_V$.
For example, a number of studies have found that the extinction law towards the Galactic centre differs significantly from the assumed `$R_V=3.1$' average \citep[e.g.][]{Udalski_only.2003,Nishiyama_Tamura.2009,Nataf_Gould.2013}.

In light of the profound impact selection effects can have on measurement of extinction, one wonders whether measurements of the shape of the extinction law could be similarly afflicted.
To investigate this we briefly consider a very simple example.
Stars were distributed in space, assigned masses and absolute magnitudes as section~\ref{sec:Ad_better_ex}, whilst extinction in the $z$ band was set to a constant value for all space.
Five different values of $A_z$ were considered, from $A_z=0.1$ to $A_z=2.0$, where $A_z \simeq 2.2 A_V$ when $R_{5495}=3.1$.
The shape of the extinction law employed was parametrized by $R_{5495}$, a monochromatic measure that is roughly equivalent to $R_V$.
The value of $R_{5495}$ for each star was drawn from a Gaussian distribution with a mean of 3.1 and a standard deviation of 0.3, following \cite{Fitzpatrick_Massa.2007}.
The extinction in each of the SDSS bands, given the values of $R_{5495}$, $A_z$ and the mass of the star were then found using the coefficients from \cite{Sale_Magorrian.2015}.
Subsequently, apparent magnitudes, including the imposition of some error, were found as in section~\ref{sec:Ad_better_ex}, with stars brighter than the magnitude limits in \textit{all} five bands included in the final catalogue.

Then the stars in the catalogue were placed in bins of equal distance modulus width (using their true distance modulus) and the mean value of $R_{5495}$ found for each bin.
The results are plotted in Fig.~\ref{fig:R5495_trend}.
When there is little extinction the retrieved mean values of $R_{5495}$ unsurprisingly match the true value.
However, as the level of extinction is increased a systematic and distance dependent bias towards larger values of $R_{5495}$ begins to appear.
The observed bias appears as smaller values of $R_{5495}$ result in more extinction in the $u$ band, for some $A_z$, therefore giving the star a fainter $u$ band apparent magnitude and so making it less likely to appear in the observed catalogue.

In light of this, it is apparent that caution should be exercised when inferring some spatial variation in the wavelength dependence of extinction, since systematic variations may appear as the result of a selection effect induced by the magnitude limits of the data employed.
Though it is worth noting that the observed effect is quite small in the presence of moderate extinction and so may be easily be overwhelmed by stochastic noise in small samples.
As with 3D extinction mapping, Poisson point processes can be used to enable the correct construction and normalisation of a likelihood function that accounts for the selection function, thus facilitating accurate inference of any extinction law variation.

\section{Closing discussion}\label{sec:close}

This paper has described how catalogues can be modelled as a realisation of a Poisson point process.
A significant benefit of this approach is that it allows the effects of selection functions to be modelled trivially, so that unbiased estimates of the parameters of astrophysical populations can be easily obtained.
Some of what has been described above is intuitive and indeed some of the results have been derived by authors who have not invoked Poisson point processes.
However, by employing Poisson point processes not only is it easier to construct likelihoods for a catalogue of observations, but it also becomes more straightforward to draw on the significant statistics and machine learning literatures that cover point processes.

Importantly, the use of a Poisson point process makes clear how to correctly normalise the likelihood of obtaining some catalogue of observations, when intuition alone might fail.
There are two distinct possibilities as to how the likelihood should be normalised, if the number of objects in the catalogue is fixed a priori then the likelihood must only be normalised over the space of observations and fundamental parameters for each object, as in section~\ref{sec:known}.
If, however, the number of objects in the catalogue is not fixed then the normalisation must cover both all possible numbers of objects as well as the space of observations and fundamental parameters for each object.
Appendix~\ref{app:norm} demonstrates that this is the nature of the normalisation term that appears in the Poisson point process likelihoods such as~\eqref{eqn:P_like} and~\eqref{eqn:like}.
Crucially, the normalisation of the likelihood typically depends on the hyperparameters, i.e. the characteristics of the population that the observed objects are drawn from.
Consequently, any approach that either ignores or incorrectly treats the normalisation of the likelihood will obtain biased/incorrect inferences about the population.

Specific focus has been given to the issue of mapping extinction in three dimensions.
This is an instance where the impact of selection functions can be particularly worrisome, as more extinguished stars will typically be fainter and so less likely to appear in any catalogue of observations.
In section~\ref{sec:Ad_better_ex} it was shown how a Poisson point process based likelihood can enable the accurate estimation of extinction as a function of distance.
This Poisson point process based approach is essentially that employed by \cite{Sale_Drew.2014} to obtain unbiased 3D maps of extinction.
Section~\ref{sec:Ad_better_ex} also examines in detail how important it is to correctly normalise the likelihoods employed, looking at the results of other formulations.

In the previous sections we have been assuming that the selection function $\selfunc$ is well known; all of the posterior distributions displayed above are conditioned on $\selfunc$.
However, in reality one will not have perfect knowledge of $\selfunc$.
Consequently, there are two fundamental options, either extend the posterior to also infer $\selfunc$ or marginalise over it.
In many cases prior knowledge of the selection function will be significant, such that marginalising over it will not introduce much extra uncertainty.
However, in some cases, including confusion limited catalogues, $\selfunc$ might not be well known and so will be an important source of uncertainty.

In section~\ref{sec:pp_cats} four assumptions were listed that must be met on order for a catalogue to be described as a realisation of a Poisson point process.
In the examples described in this paper all of these assumptions are met.
However, there are other instances where this will not be the case.
Specifically, the method of \cite{Sale_Magorrian.2014} fails to meet condition~\ref{list:1}:
as the extinction to the stars in the catalogue is modelled using a Gaussian random field, the extinction to two stars is explicitly covariant and so the assumption of independence is no longer valid.
An extension to the approach described here, that invokes Markov point processes and can then be used with the method of \cite{Sale_Magorrian.2014}, will be the subject of a future paper (Sale in prep.).

\section*{Acknowledgements}

I would like to thank John Magorrian for his comments and suggestions that have improved this paper.

The research leading to the results presented here was supported by
the United Kingdom Science Technology and Facilities Council (STFC,
ST/K00106X/1) and the European Research Council under the European
Union’s Seventh Framework Programme (FP7/2007-2013)/ERC grant
agreement no. 321067.

\bibliography{astroph_3,bibliography-2,ml,temp}

\newcommand{\noop}[1]{}
\begin{thebibliography}{46}
\expandafter\ifx\csname natexlab\endcsname\relax\def\natexlab#1{#1}\fi

\bibitem[{Adams, Murray \& MacKay(2009)Adams, Murray, \&
  MacKay}]{Adams_Murray.2009}
Adams R.~P., Murray I., MacKay D. J.~C., 2009, in Proceedings of the 26th
  International Conference on Machine Learning (ICML), Bottou L., Littman M.,
  eds., Omnipress, Montreal, pp. 9--16

\bibitem[{{Berry} {et~al}\mbox{.}(2012){Berry}, {Ivezi{\'c}}, {Sesar},
  {Juri{\'c}}, {Schlafly}, {Bellovary}, {Finkbeiner}, {Vrbanec}, {Beers},
  {Brooks}, {Schneider}, {Gibson}, {Kimball}, {Jones}, {Yoachim}, {Krughoff},
  {Connolly}, {Loebman}, {Bond}, {Schlegel}, {Dalcanton}, {Yanny}, {Majewski},
  {Knapp}, {Gunn}, {Allyn Smith}, {Fukugita}, {Kent}, {Barentine},
  {Krzesinski}, \& {Long}}]{Berry_Ivezic.2012}
{Berry} M. {et~al.}, 2012, \apj, 757, 166

\bibitem[{Diggle(2003)}]{Diggle_only.2003}
Diggle P., 2003, Statistical analysis of spatial point patterns. Edward Arnold,
  2nd edition

\bibitem[{Doucet, De~Freitas \& Gordon(2001)Doucet, De~Freitas, \&
  Gordon}]{Doucet_DeFreitas.2001}
Doucet A., De~Freitas N., Gordon N., 2001, An introduction to sequential Monte
  Carlo methods. Springer

\bibitem[{{Eddington}(1913)}]{Eddington_only.1913}
{Eddington} A.~S., 1913, \mnras, 73, 359

\bibitem[{Erlang(1909)}]{Erlang_only.1909}
Erlang A.~K., 1909, Nyt Tidsskrift for Matematik B, 20, 16

\bibitem[{{Fischera} \& {Dopita}(2004)}]{Fischera_Dopita.2004}
{Fischera} J., {Dopita} M.~A., 2004, \apj, 611, 919

\bibitem[{{Fitzgerald}(1968)}]{Fitzgerald_only.1968}
{Fitzgerald} M.~P., 1968, \aj, 73, 983

\bibitem[{{Fitzpatrick}(2004)}]{Fitzpatrick_only.2004}
{Fitzpatrick} E.~L., 2004, in Astronomical Society of the Pacific Conference
  Series, Vol. 309, Astrophysics of Dust, {Witt} A.~N., {Clayton} G.~C.,
  {Draine} B.~T., eds., p.~33

\bibitem[{{Fitzpatrick} \& {Massa}(2007)}]{Fitzpatrick_Massa.2007}
{Fitzpatrick} E.~L., {Massa} D., 2007, \apj, 663, 320

\bibitem[{{Foreman-Mackey}, {Hogg} \& {Morton}(2014){Foreman-Mackey}, {Hogg},
  \& {Morton}}]{Foreman-Mackey_Hogg.2014}
{Foreman-Mackey} D., {Hogg} D.~W., {Morton} T.~D., 2014, \apj, 795, 64

\bibitem[{{Gajjar}, {Joshi} \& {Kramer}(2012){Gajjar}, {Joshi}, \&
  {Kramer}}]{Gajjar_Joshi.2012}
{Gajjar} V., {Joshi} B.~C., {Kramer} M., 2012, \mnras, 424, 1197

\bibitem[{Geiger, Verma \& Pearl(1990)Geiger, Verma, \&
  Pearl}]{Geiger_Verma.1990}
Geiger D., Verma T., Pearl J., 1990, Networks, 20, 507

\bibitem[{{Green} {et~al}\mbox{.}(2014){Green}, {Schlafly}, {Finkbeiner},
  {Juri{\'c}}, {Rix}, {Burgett}, {Chambers}, {Draper}, {Flewelling},
  {Kudritzki}, {Magnier}, {Martin}, {Metcalfe}, {Tonry}, {Wainscoat}, \&
  {Waters}}]{Green_Schlafly.2014}
{Green} G.~M. {et~al.}, 2014, \apj, 783, 114

\bibitem[{{Hanson} \& {Bailer-Jones}(2014)}]{Hanson_Bailer-Jones.2014}
{Hanson} R.~J., {Bailer-Jones} C.~A.~L., 2014, \mnras, 438, 2938

\bibitem[{Illian {et~al}\mbox{.}(2008)Illian, Penttinen, Stoyan, \&
  Stoyan}]{Illian_Penttinen.2008}
Illian J., Penttinen A., Stoyan H., Stoyan D., 2008, Statistical analysis and
  modelling of spatial point patterns. John Wiley \& Sons

\bibitem[{{Jeffreys}(1938)}]{Jeffreys_only.1938}
{Jeffreys} H., 1938, \mnras, 98, 190

\bibitem[{Kingman(1992)}]{Kingman_only.1992}
Kingman J. F.~C., 1992, Poisson processes, Vol.~3. Oxford university press

\bibitem[{{Lallement} {et~al}\mbox{.}(2014){Lallement}, {Vergely}, {Valette},
  {Puspitarini}, {Eyer}, \& {Casagrande}}]{Lallement_Vergely.2014}
{Lallement} R., {Vergely} J.-L., {Valette} B., {Puspitarini} L., {Eyer} L.,
  {Casagrande} L., 2014, \aap, 561, A91

\bibitem[{{Lombardi}, {Lada} \& {Alves}(2013){Lombardi}, {Lada}, \&
  {Alves}}]{Lombardi_Lada.2013}
{Lombardi} M., {Lada} C.~J., {Alves} J., 2013, \aap, 559, A90

\bibitem[{{Loredo}(2004)}]{Loredo_only.2004}
{Loredo} T.~J., 2004, in American Institute of Physics Conference Series, Vol.
  735, American Institute of Physics Conference Series, {Fischer} R., {Preuss}
  R., {Toussaint} U.~V., eds., pp. 195--206

\bibitem[{{Lutz} \& {Kelker}(1973)}]{Lutz_Kelker.1973}
{Lutz} T.~E., {Kelker} D.~H., 1973, \pasp, 85, 573

\bibitem[{{Majewski}, {Zasowski} \& {Nidever}(2011){Majewski}, {Zasowski}, \&
  {Nidever}}]{Majewski_Zasowski.2011}
{Majewski} S.~R., {Zasowski} G., {Nidever} D.~L., 2011, \apj, 739, 25

\bibitem[{{Malmquist} \& {Hufnagel}(1933)}]{Malmquist_Hufnagel.1933}
{Malmquist} K.~G., {Hufnagel} L., 1933, Stockholms Observatoriums Annaler, 11,
  9

\bibitem[{{Marigo} {et~al}\mbox{.}(2008){Marigo}, {Girardi}, {Bressan},
  {Groenewegen}, {Silva}, \& {Granato}}]{Marigo_Girardi.2008}
{Marigo} P., {Girardi} L., {Bressan} A., {Groenewegen} M.~A.~T., {Silva} L.,
  {Granato} G.~L., 2008, \aap, 482, 883

\bibitem[{{Marshall} {et~al}\mbox{.}(2006){Marshall}, {Robin}, {Reyl{\'e}},
  {Schultheis}, \& {Picaud}}]{Marshall_Robin.2006}
{Marshall} D.~J., {Robin} A.~C., {Reyl{\'e}} C., {Schultheis} M., {Picaud} S.,
  2006, \aap, 453, 635

\bibitem[{Moller \& Waagepetersen(2004)}]{Moller_Waagepetersen.2004}
Moller J., Waagepetersen R.~P., 2004, Statistical inference and simulation for
  spatial point processes. CRC Press

\bibitem[{{Nataf} {et~al}\mbox{.}(2013){Nataf}, {Gould}, {Fouqu{\'e}},
  {Gonzalez}, {Johnson}, {Skowron}, {Udalski}, {Szyma{\'n}ski}, {Kubiak},
  {Pietrzy{\'n}ski}, {Soszy{\'n}ski}, {Ulaczyk}, {Wyrzykowski}, \&
  {Poleski}}]{Nataf_Gould.2013}
{Nataf} D.~M. {et~al.}, 2013, \apj, 769, 88

\bibitem[{{Neckel}(1966)}]{Neckel_only.1966}
{Neckel} T., 1966, Zeitschrift fur Astrophysik, 63, 221

\bibitem[{{Neckel} \& {Klare}(1980)}]{Neckel_Klare.1980}
{Neckel} T., {Klare} G., 1980, \aaps, 42, 251

\bibitem[{{Nishiyama} {et~al}\mbox{.}(2009){Nishiyama}, {Tamura}, {Hatano},
  {Kato}, {Tanab{\'e}}, {Sugitani}, \& {Nagata}}]{Nishiyama_Tamura.2009}
{Nishiyama} S., {Tamura} M., {Hatano} H., {Kato} D., {Tanab{\'e}} T.,
  {Sugitani} K., {Nagata} T., 2009, \apj, 696, 1407

\bibitem[{{Ostriker}, {Stone} \& {Gammie}(2001){Ostriker}, {Stone}, \&
  {Gammie}}]{Ostriker_Stone.2001}
{Ostriker} E.~C., {Stone} J.~M., {Gammie} C.~F., 2001, \apj, 546, 980

\bibitem[{Pearl(1988)}]{Pearl_only.1988}
Pearl J., 1988, Probabilistic reasoning in intelligent systems: Networks of
  plausible inference. Morgan Kaufmann

\bibitem[{{Sale}(2012)}]{Sale_only.2012}
{Sale} S.~E., 2012, \mnras, 427, 2119

\bibitem[{{Sale} {et~al}\mbox{.}(2014){Sale}, {Drew}, {Barentsen}, {Farnhill},
  {Raddi}, {Barlow}, {Eisl{\"o}ffel}, {Vink}, {Rodr{\'{\i}}guez-Gil}, \&
  {Wright}}]{Sale_Drew.2014}
{Sale} S.~E. {et~al.}, 2014, \mnras, 443, 2907

\bibitem[{{Sale} {et~al}\mbox{.}(2009){Sale}, {Drew}, {Unruh}, {Irwin},
  {Knigge}, {Phillipps}, {Zijlstra}, {G{\"a}nsicke}, {Greimel}, {Groot},
  {Mampaso}, {Morris}, {Napiwotzki}, {Steeghs}, \& {Walton}}]{Sale_Drew.2009}
---, 2009, \mnras, 392, 497

\bibitem[{{Sale} \& {Magorrian}(2014)}]{Sale_Magorrian.2014}
{Sale} S.~E., {Magorrian} J., 2014, \mnras, 445, 256

\bibitem[{{Sale} \& {Magorrian}(2015)}]{Sale_Magorrian.2015}
---, 2015, \mnras, 448, 1738

\bibitem[{{Sanders} \& {Binney}(2015)}]{Sanders_Binney.2015}
{Sanders} J.~L., {Binney} J., 2015, \mnras, 449, 3479

\bibitem[{{Sarazin}(1980)}]{Sarazin_only.1980}
{Sarazin} C.~L., 1980, \apj, 236, 75

\bibitem[{{Schultheis} {et~al}\mbox{.}(2014){Schultheis}, {Chen}, {Jiang},
  {Gonzalez}, {Enokiya}, {Fukui}, {Torii}, {Rejkuba}, \&
  {Minniti}}]{Schultheis_Chen.2014}
{Schultheis} M. {et~al.}, 2014, \aap, 566, A120

\bibitem[{{Tempel} {et~al}\mbox{.}(2014){Tempel}, {Stoica}, {Mart{\'{\i}}nez},
  {Liivam{\"a}gi}, {Castellan}, \& {Saar}}]{Tempel_Stoica.2014}
{Tempel} E., {Stoica} R.~S., {Martinez} V.~J., {Liivam{\"a}gi} L.~J.,
  {Castellan} G., {Saar} E., 2014, \mnras, 438, 3465

\bibitem[{{Tierney}(1994)}]{Tierney_only.1994}
{Tierney} L., 1994, Annals of Statistics, 22, 1701

\bibitem[{{Udalski}(2003)}]{Udalski_only.2003}
{Udalski} A., 2003, \apj, 590, 284

\bibitem[{{Vergely} {et~al}\mbox{.}(2010){Vergely}, {Valette}, {Lallement}, \&
  {Raimond}}]{Vergely_Valette.2010}
{Vergely} J.-L., {Valette} B., {Lallement} R., {Raimond} S., 2010, \aap, 518,
  A31

\bibitem[{{Zasowski} {et~al}\mbox{.}(2013){Zasowski}, {Johnson}, {Frinchaboy},
  {Majewski}, {Nidever}, {Rocha Pinto}, {Girardi}, {Andrews}, {Chojnowski},
  {Cudworth}, {Jackson}, {Munn}, {Skrutskie}, {Beaton}, {Blake}, {Covey},
  {Deshpande}, {Epstein}, {Fabbian}, {Fleming}, {Garcia Hernandez}, {Herrero},
  {Mahadevan}, {M{\'e}sz{\'a}ros}, {Schultheis}, {Sellgren}, {Terrien}, {van
  Saders}, {Allende Prieto}, {Bizyaev}, {Burton}, {Cunha}, {da Costa},
  {Hasselquist}, {Hearty}, {Holtzman}, {Garc{\'{\i}}a P{\'e}rez}, {Maia},
  {O'Connell}, {O'Donnell}, {Pinsonneault}, {Santiago}, {Schiavon}, {Shetrone},
  {Smith}, \& {Wilson}}]{Zasowski_Johnson.2013}
{Zasowski} G. {et~al.}, 2013, \aj, 146, 81

\end{thebibliography}

\appendix

\section{Marginal and conditional independence}\label{app:ind}

If $A$ and $B$ are conditionally independent then
\begin{align}
\pr(A,B|C) &= \pr(A|C)\pr(B|C) \\
\pr(A|B,C) &= \pr(A|C) \\
\pr(B|A,C) &= \pr(B|C)
\end{align}
On the other hand, if $A$ and $B$ are marginally independent then
\begin{align}
\pr(A,B) &= \pr(A)\pr(B) \\
\pr(A|B) &= \pr(A) \\
\pr(B|A) &= \pr(B)
\end{align}
This state is also often referred to simply as independence, though the use of the term marginal independence is preferred here to reduce ambiguity.
The origin of the nomenclature is more obvious if we consider that the marginal independence of $A$ and $B$ also implies, e.g.
\begin{equation}
\begin{split}
\pr(A|B) &= \int \pr(A,C | B)\, dC \\
&=  \pr(A) ,
\end{split}
\end{equation}
where
\begin{equation}
\pr(A,C|B) \neq \pr(A,C).
\end{equation}
Note that marginal independence does not imply conditional independence, nor does conditional independence imply marginal independence.

\section{Normalisation factor $Z(\hyper)$}\label{app:norm}

The exponential term in~\eqref{eqn:P_like} is simply a normalisation factor, covering all possible number of observed points, from 0 to $\infty$, as well as the entire space $\mathcal{Y}$ for each point.
In order to demonstrate this, we start by defining the likelihood,
\begin{equation}
\pr(\data | \hyper) = \frac{1}{Z(\hyper) } \prod_{n=0}^{\ntot} \lambda(\vy_n|\hyper).
\end{equation}
We can begin constructing the normalisation $Z(\hyper)$ by considering the normalisation for a fixed number of points $n$
\begin{equation}
Z(\hyper|n) = \frac{1}{n!} \int_{\mathcal{Y}} \cdots \int_{\mathcal{Y}} \prod_{n=0}^{n} \lambda(\vy_n|\hyper) \, dy_1 \ldots dy_n ,
\end{equation}
where the factorial appears as we can not discriminate between different orderings of the points.
Then if we consider all values of $n$
\begin{equation}
\begin{split}
Z(\hyper) &= \sum_{n=0}^{\infty} \frac{1}{n!} \int_{\mathcal{Y}} \cdots \int_{\mathcal{Y}} \prod_{n=1}^{n} \lambda(\vy_n|\hyper) \, dy_1 \ldots dy_n \\
& = \sum_{n=0}^{\infty} \frac{1}{n!} \left(\int_{\mathcal{Y}} \lambda(\vy|\hyper) \, dy \right)^n \\
& = \exp \left( \int_{\mathcal{Y}} \lambda(\vy|\hyper) \, dy \right) ,
\end{split}
\end{equation}
where this is the normalisation factor in ~\eqref{eqn:P_like}.
This short demonstration follows from e.g. \cite{Moller_Waagepetersen.2004}.

\section{Details of Malmquist bias example}\label{app:Malm_ex}

\begin{figure}
\includegraphics{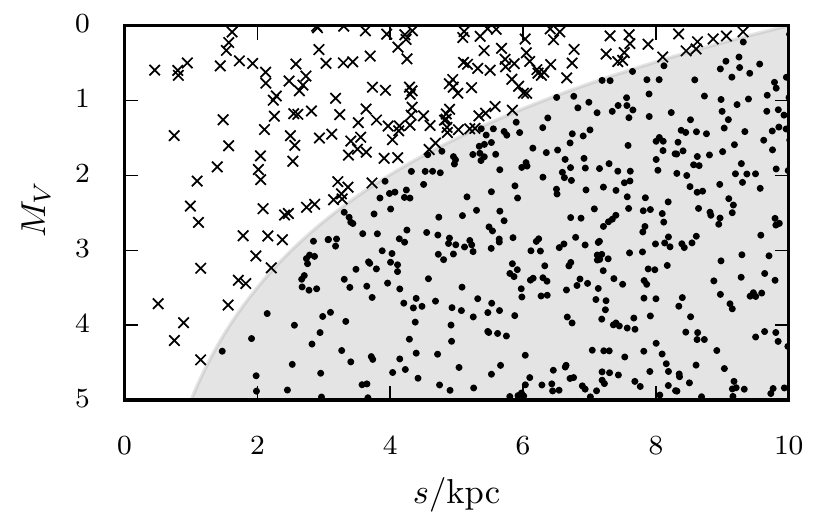}
\caption{The catalogue represented in distance--absolute magnitude space. Crosses represent stars that appear in the catalogue, points represent those that are too faint and so are not included in the catalogue.
  \label{fig:sim_cat}}
\end{figure}

In section~\ref{sec:Malm_ex} an example that demonstrated the use of Poisson point process based inference of the spatial distribution of stars in the presence of Malmquist bias was briefly discussed.
We will now discuss more details of this example.

The catalogue studied consists of a distance and a $V$-band apparent magnitude for each star and was simulated subject to a short list of assumptions. 
The number density of the observed stars is assumed to decline exponentially with distance, with the rate of decline set by a scale length of 3~kpc.
The stars all exist on a square region of the sky, of solid angle $\Omega$, as defined by e.g. RA and DEC.
The luminosity function of the stars is assumed to be uniform across the range $0 \leq V < 5$ and the absolute magnitude of the stars is assumed to be marginally independent of their distance, with
\begin{equation}
\pr(M_V) = \begin{cases} \frac{1}{5} &\mbox{if } 0 \leq M_V < 5; \\
0 &\mbox{otherwise. } \end{cases}  \label{eqn:Malm_LF}
\end{equation}
at all distances.
The catalogue is subject to a step function faint magnitude limit at $V=15$: all objects with apparent magnitudes brighter than this appear in the catalogue, whilst those that are fainter do not.
Finally we assume the distance and absolute magnitude of each star are measured with absolute precision and effects such as interstellar extinction, binarity and so on are not included.
The simulated catalogue~$\data_M$ is shown in Fig.~\ref{fig:sim_cat}.

Given $\data_M$ we now seek to infer the scale length of the population utilising the Poisson point process based methodology.
We can first define the probability distribution for the distance $s_n$ to the $n{\rm th}$ star as
\begin{equation}
\pr(s_n | L) = \frac {s_n^2 e^{-s_n/L}}{2L^3}, \label{eqn:Malm_dist}
\end{equation}
where $L$ is the scale length we wish to infer and the factor of $s_n^2$ arises from the Jacobian converting from Cartesian to spherical polar coordinates.
The selection function is
\begin{equation}
\selfunc(V) = \begin{cases} 1 &\mbox{if } V \leq 15 ;\\
0 &\mbox{otherwise. } \end{cases}  \label{eqn:Malm_sel}
\end{equation} 
Here the apparent magnitude $V$ is defined normally, i.e. 
\begin{equation}
V=M_V+5\log_{10}(s/10~{\rm pc}).
\end{equation}
Collectively~\eqref{eqn:Malm_LF},~\eqref{eqn:Malm_dist} and~\eqref{eqn:Malm_sel} define the intensity function, so that
\begin{equation}
\lambda(s, M_V | \norm, L, \selfunc) = \begin{cases} \frac {\norm s_n^2 e^{-s_n/L}}{10L^3} &\mbox{if } V \leq 15, \; 0 \leq M_V < 5; \\
0 &\mbox{otherwise, } \end{cases} \label{eqn:Malm_intensity}
\end{equation} 
where $\norm$ has again been introduced to indicate the total number of stars, both those in the catalogue and those that fail to make the cut.

As the number of stars in the catalogue is not known a priori, but depends on the realisation of the underlying Poisson point process, we follow the treatment discussed in section~\ref{sec:prior_post}.
If we assume uniform and independent priors on both the scale length and the total number of stars, we have the posterior distribution
\begin{equation}
\begin{split}
\pr(\norm, L | \data_M, \selfunc) \propto& \exp( - \int \int ds dM_V \lambda(s, M_V | \norm, L, \selfunc) )\\
&\times \prod_{n=1}^{\ntot} \lambda(s_n, M_{V,i} | \norm, L, \selfunc) .
\end{split}
\end{equation}
A simple Metropolis MCMC sampler was employed to explore this posterior.

\section{The methodology of Green et al. (2014)}\label{app:green}

We will now consider in detail why the method of \cite{Green_Schlafly.2014} struggles to obtain an accurate estimate of the true extinction as a function of distance, as shown in Fig.~\ref{fig:extn_better}.
Although we will not discuss it in detail, the simple \cite{Sale_only.2012} based result also performs poorly, owing to the fact that it largely fails to present a proper treatment for the selection function.

In addition to neglecting differential extinction there are a two key differences between the posterior distributions used by \cite{Green_Schlafly.2014} and those described here and applied in \cite{Sale_Drew.2014}.
These specifically focus on how the likelihood is normalised.
Using the notation adopted here, their likelihood is, as given by their equation (5), 
\begin{equation}
\pr(\data | \hyper) = \prod_{n=1}^{\ntot} \pr(\vy_n | \hyper) .
\end{equation}
By comparison to~\eqref{eqn:like_known} we can see that, if $\pr(\vy_n | \hyper)$ is properly normalised, the right hand side would be the correct likelihood if and only if the number of stars to be observed was fixed.
However, in that case, the likelihood would be conditioned on $\ntot$ so that the left hand side would instead read $\pr(\data | \ntot, \hyper)$.
\cite{Green_Schlafly.2014} appear to intend that their method is applied to photometry, in which case the number of stars in the observed catalogue is a priori unknown and so they require a likelihood \textit{not} conditioned on the size of the catalogue.
In that case they do not have the correct normalisation as they have not performed the normalisation over the a priori unknown number of stars in the catalogue, that gives rise to the exponential term in~\eqref{eqn:P_like} that is missing in their formulation (see appendix~\ref{app:norm}).
Though, this problem need not be completely fatal.
As discussed in section~\ref{sec:prior_post}, a judicious choice of $\pr(\norm)$, albeit a $\pr(\norm)$ that is a posteriori unlikely, can enable one to arrive at a posterior distribution similar to what \cite{Green_Schlafly.2014} reach.

\begin{figure}
\includegraphics{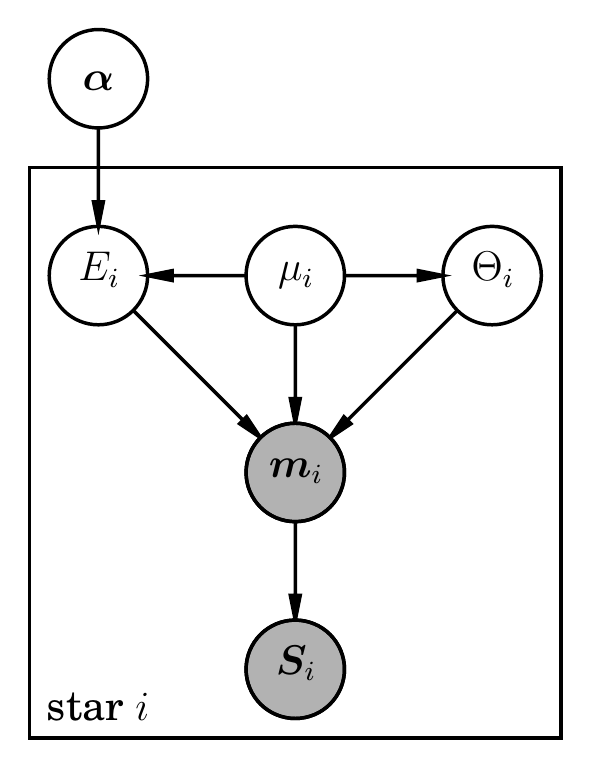}
\caption{A directed acyclic graph that illustrates the statistical model adopted by \protect\cite{Green_Schlafly.2014}.
  \label{fig:Greeen_graph}}
\end{figure}

However, it is also apparent that their definition of $\pr(\vy_n | \hyper)$, or in their notation $\pr(\vm | \bm{\alpha}, \vS)$, is itself not correctly normalised, since, as can be seen in ~\eqref{eqn:like_known}, the normalisation factor should depend on the hyperparameters.
The source of this error in the normalisation is also relatively subtle.
Adopting their notation, \cite{Green_Schlafly.2014} essentially write that, for each star
\begin{equation}
\begin{split}
\pr(\vm | \bm{\alpha}) &= \iiint \, d\mu\,dE\,d\Theta\, \pr(\vm, \mu, E, \Theta | \bm{\alpha}) \\
&= \iiint \, d\mu\,dE\,d\Theta\, \pr(\vm | \mu, E, \Theta) \pr( \mu, E, \Theta| \bm{\alpha}) \\
&= \iiint \, d\mu\,dE\,d\Theta\, \pr(\vm | \mu, E, \Theta) \pr(E | \mu, \bm{\alpha}) \pr(\mu,\Theta) . \label{eqn:GS_star_like1}
\end{split}
\end{equation}
Where $\vm$ represent the observed apparent magnitudes of the star, $\mu$ its distance modulus, $E$ the extinction to the star, $\Theta$ the star's spectral type and $\bm{\alpha}$ are the hyperparameters that control the distance extinction relationship.
$\pr(E | \mu, \bm{\alpha})$ takes the form of a $\delta$-function, non-zero at an extinction defined by $\bm{\alpha}$ for each $\mu$, thus making it straightforward to integrate over $E$.
They then note that the above integrand is proportional to the posterior $\pr(\mu, E, \Theta | \vm, \rm{flat})$, that is derived when placing a flat prior on $E$.
So that
\begin{equation}
\pr(\vm | \bm{\alpha}) = \pr(\vm|\rm{flat}) \iiint \, d\mu\,dE\,d\Theta\, \pr(\mu, E, \Theta | \vm, \rm{flat}) \frac{\pr(E | \mu, \bm{\alpha})}{\pr(E| \rm{flat})} . \label{eqn:GS_star_like_2}
\end{equation}

They then go on to introduce $\vS$, a vector that indicates whether the star has been observed in each photometric band, and define the posterior distribution $\pr(\mu, E, \Theta | \vm, \vS, \rm{flat})$ in their equations~(34) to~(40).
They then assert that it is possible to simply employ this in place of $\pr(\mu, E, \Theta | \vm, \rm{flat})$, as in~\label{eqn:GS_star_like2}, in order to find $\pr(\vm | \bm{\alpha}, \vS)$, but 
\begin{equation}
\pr(\vm | \bm{\alpha}, \vS) \neq \pr(\vm | \vS, \rm{flat}) \iiint \, d\mu\,dE\,d\Theta\, \pr(\mu, E, \Theta | \vm, \vS, \rm{flat}) \frac{\pr(E | \mu, \bm{\alpha})}{\pr(E | \rm{flat})} . 
\end{equation}

The derivation in~\eqref{eqn:GS_star_like1} and~\eqref{eqn:GS_star_like_2} relied on the fact that 
\begin{equation}
\pr( \mu, E, \Theta| \bm{\alpha}) = \pr(E | \mu, \bm{\alpha}) \pr(\mu,\Theta) \label{eqn:GS_prior_split} .
\end{equation}
However, this does not carry over when conditioning on the star being observed (or not) in each band, i.e.
\begin{equation}
\pr( \mu, E, \Theta| \bm{\alpha}, \vS) \neq \pr(E | \mu, \bm{\alpha}, \vS) \pr(\mu,\Theta| \vS).\label{eqn:GS_prior_split_cond}
\end{equation}
The reason for this can be seen by considering a graph of the model they propose (Fig.~\ref{fig:Greeen_graph}) and the rules of d-separation \citep{Pearl_only.1988, Geiger_Verma.1990}.
Specifically, in order for $E$ to be independent of $\Theta$, the apparent magnitudes $\vm$ and its children must not appear in the conditioning set.
In~\eqref{eqn:GS_prior_split} this requirement is met.
However, once conditioned upon $\vS$, as in \eqref{eqn:GS_prior_split_cond}, this is no longer the case, since $\vS$ is a child of $\vm$.

Fortunately, it is possible to remedy this problem.
A sensible starting point for considering conditioning the likelihood on the fact that the data have been observed is to note that
\begin{equation}
\pr(\vm | \bm{\alpha}, \vS) = \frac{\pr(\vm, \vS | \bm{\alpha})}{\pr(\vS | \bm{\alpha})}.
\end{equation}
where
\begin{equation}
\begin{split}
\pr(\vm, \vS | \bm{\alpha}) &= \iiint \, d\mu\,dE\,d\Theta\, \pr(\vm, \vS | \mu, E, \Theta) \pr(E | \mu, \bm{\alpha}) \pr(\mu,\Theta)  \\
&= \pr(\vm, \vS | \rm{flat}) \iiint \, d\mu\,dE\,d\Theta\, \pr(\mu, E, \Theta | \vm, \vS, \rm{flat}) \frac{\pr(E | \mu, \bm{\alpha})}{\pr(E| \rm{flat})} , 
\end{split}
\end{equation}
where $\pr(\mu, E, \Theta | \vm, \vS, \rm{flat})$ is given by their~(36).
Meanwhile,
\begin{equation}
\pr(\vS | \bm{\alpha}) = \iiiint \, d\mu\,dE\,d\Theta\,d\vm\, \pr(\vS|\vm) \pr(\vm, \vS | \mu, E, \Theta) \pr(E | \mu, \bm{\alpha}) .
\end{equation}
This is the average probability of any given star being observed subject to some distance extinction relationship as parametrized by $\bm{\alpha}$.
If extinction quickly rises to large values this probability will be small, whilst if extinction is small at all distances this probability will be much greater.
As a consequence, the effect of this term will be to push the inferred distance-extinction relationship towards values the selection function might normally prevent it reaching.   
So, for example, it will account for the more extinguished stars that would have been observed if only they were not too faint.

\end{document}